\documentclass[twocolumn,showpacs,preprintnumbers,amsmath,amssymb,nofootinbib,pra,superscriptaddress,10pt]{revtex4-1}
\usepackage{graphicx}% Include figure files
\usepackage{dcolumn}% Align table columns on decimal point
\usepackage{bm}% bold math
\usepackage{epstopdf} %% added by claudio

\usepackage{amssymb} %vari simboli matematici
\usepackage{amsmath} %vari simboli matematici
\usepackage{latexsym} %vari simboli matematici
\usepackage{amsthm} %per i teoremi
\usepackage{mathrsfs} %vari simboli matematici

\theoremstyle{plain}
\newtheorem{theo}{Theorem}[section]

\theoremstyle{definition}

\theoremstyle{remark}

\begin{document}

%\preprint{APS/123-QED}

\title{Irreversible Evolution of a Wave Packet in The Rigged Hilbert Space Quantum Mechanics.}

\author{Giulia Marcucci}
   \affiliation{Institute for Complex Systems, National Research Council (ISC-CNR), Via dei Taurini 19, 00185 Rome (IT).}
    \email{marcucci.giulia@gmail.com}
\author{Claudio Conti}
    \affiliation{Institute for Complex Systems, National Research Council (ISC-CNR), Via dei Taurini 19, 00185 Rome (IT).}
      \affiliation{Department of Physics, University Sapienza, Piazzale Aldo Moro 5, 00185 Rome (IT).}
    \homepage{http://www.complexlight.org}

\date{\today}

\begin{abstract}
It is well known that a state with complex energy cannot be the eigenstate of a self-adjoint operator, as the Hamiltonian. Resonances, i.e. states with exponentially decaying observables, are not vectors belonging to the conventional Hilbert space. One can describe these resonances in an unusual mathematical formalism, based on the so-called Rigged Hilbert Space (RHS). In the RHS, the states with complex energy are denoted as Gamow Vectors (GV), and they model decay processes.

We study GV of the Reversed Harmonic Oscillator (RHO), and we analytically and numerically investigate the unstable evolution of wave packets. We introduce the background function to study initial data not composed only by a summation of GV and we analyse different wave packets belonging to specific function spaces. Our work furnishes support to the idea that irreversible wave propagations can be investigated using Rigged Hilbert Space Quantum Mechanics and provides insights for the experimental investigation of irreversible dynamics.
\end{abstract}

\maketitle

\section{\label{sec:intro}Introduction}
The spontaneous decay of nuclear particles, or the empirical evidences for the Big Bang, lead various authors \cite{bohm1, bohm2, bohm3, prigogine, antoniou,prigogine} to consider modifications of the principles of quantum mechanics in order to include time-asymmetry. Despite a long standing effort, the need for these modifications is still debated. However, the debate stimulated the developments of theoretical tools and paradigms by a growing community of scientists. These tools recently found surprisingly applications in nonlinear physics and photonics \cite{noi1, noi2, longhi}.

The time-asimmetric dynamics of quantum systems are also relevant in biophysics, fluid dynamics, network theory, entanglement generation and epigenetic studies (see for example \cite{materassi, boccaletti, calvani} and the references therein). The decay from local maxima in the energetic landscape of complex systems is generically retained irreversible. However the origin of this irreversibility in the case of microscopic quantum structures, or in quantum inspired models of networks \cite{perra}, is unconsidered so far \cite{liu}.

The leading theoretical background of Time-Asymmetric Quantum Mechanics (TA-QM) is the RHS, an enlarged Hilbert space, which includes non-normalizable wave packets that get amplified, or decay exponentially with time. The paradigmatic model for TA-QM is the reversed harmonic oscillator (RHO). Whithin standard QM the description of the evolution of a wavefunction may be done by using the continuous spectrum of the Hamiltonian.
In TA-QM one considers a RHS, and generalized eigenvalues with complex energies do have physical meanings \cite{aglietti}. The corresponding non-normalizable eigenvectors are the so-called ``Gamow vectors'' \cite{gamow1, gamow2} and form a numerable generalized basis for integrable functions.
This discrete resummation of the continuous spectrum furnishes novel physical insights for the spontaneous decay of a wave packet.
For example, one can predict the surprising result of the quantization of the decay rates, which has been experimentally observed in an optical emulation \cite{noi2,mari}.

A further intriguing outcome of the TA-QM is the fact that it unveils a particular structure in the phase space. Namely, one can discriminate initial data in terms of their projection on the eigenvectors of the RHO continuous spectrum. These projections engender different vector subspaces depending on specific Hardy spaces. The question if this classification has direct physical counterpart has no answer at the moment,
albeit mathematically this has important implications. If the initial wavefunction belongs to a particular space, it is completely represented by a numerable set of generalized eigenfunctions. On the contrary, the representation is given in terms of a finite number of GV and a background function.

Despite these mathematical properties have been studied by several authors \cite{chruscinski1, chruscinski2, bollini, civitarese, delamadrid, gadella,barton, celeghini}, a direct physical evidence of their implications is lacking, even in the simplest case of RHO. 
In this manuscript, we review the basics of TA-QM and of the GV approach to the RHO. Moreover, we study the way the function space of the input wavefunctions has a direct counterpart in the long term evolution. 

This article is organized as follows: in sec. \ref{sec:ths} we present three theorems which establish that quantum mechanics with a temporal asymmetry cannot exist in the standard Hilbert space formulation. In sec. \ref{sec:rhs} we build a different topology for the space of initial data and show the way this new geometry causes an enlargement of the Hilbert space, namely, the rigged Hilbert space. Direct consequences of RHS are disclosed in sec. \ref{sec:eigen}. In sec. \ref{sec:damped} we describe quantized damped motion, and in sec. \ref{sec:rho} we study RHO. In sec. \ref{sec:compact} and \ref{sec:gaussian} we study the evolution of a function with compact support and a Gaussian function, respectively. We show the differences between their propagation: since the first one belongs to a specific function space, it exponentially decays with time; on the contrary, since the Gaussian function does not have a compact support, it is the superposition of exponentially and algebraically decaying waves. Conclusions are drawn in sec. \ref{sec:conclusions}. 

\section{\label{sec:ths}Fundamental theorems of QM}
In order to build a mathematical theory behind a generic quantum system we need to define a Hausdorff vector space $\Psi$, with a locally convex topology $\tau$ and a scalar product $(\cdot|\cdot)$. We need also an algebra $\mathcal{A}$ of $\tau$-continuous linear operator on $\Psi$ and a probability measure $\mathcal{P}$ on $\mathcal{A}$.
By the scalar product $(\cdot|\cdot)$, we are able to build a norm $||\psi||=\sqrt{(\psi|\psi)}$ $\forall \psi \in \Psi$ and a metric $d(\psi,\phi)=||\phi-\psi||$ $\forall\,\phi,\psi\in\Psi$, that is induced by the norm, therefore we can settle a new topology $\tau_d$ on $\Psi$, given by the distance $d$. Now, we have a Euclidean space $(\Psi,\tau_d)$, which is also normed and separable. To be a physical space needs the completeness.

Let $(\mathcal{H},\tau_{\mathcal{H}})$ be the completion of $(\Psi,\tau_d)$; $\mathcal{H}$ is a separable Hilbert space, and is the space used to formulate the known \textit{time symmetric} quantum theory. The temporal simmetry in a Hilbert space arises from the following three theorems:

\begin{theo}[Gleason]
\label{gleason}
\cite{gleason}
For every probability $\mathcal{P}(\Lambda)$, there exists a positive trace class operator $\rho$ such that
$$\mathcal{P}(\Lambda)=Tr(\Lambda\rho).$$
\end{theo}

\begin{theo}[Stone-Neumann]
\label{stone-neumann}
\cite{stone}
Let us consider the Schr\"odinger-Neumann equation for $\rho$ previously defined
$$\frac{\partial\rho(t)}{\partial t}=\frac{i}{\hbar}[H,\rho(t)],$$
with $H$ Hamiltonian operator.
The solutions of such an equation are time simmetric and they are given by the group of unitary operators $U^{\dagger}(t)=\exp{-\frac{i}{\hbar}Ht}$.
\end{theo}

\begin{theo}[Hegerfeldt]
\label{hegerfeldt}
\cite{hegerfeldt}
For every Hermitian and semi-bounded Hamiltonian $H$, either
$$Tr(\Lambda(t)\rho)=Tr(\Lambda\rho(t))=0 \;\; \forall t\in\mathbb{R}$$
or
$$Tr(\Lambda(t)\rho)=Tr(\Lambda\rho(t))>0 \;\; \forall t\in\mathbb{R}$$
except on a set of Lebesgue measure zero.
\end{theo}

These theorems imply that time asymmetric solutions of the Schr\"odinger equation
$$i\hbar\frac{\partial\phi(t)}{\partial t}=H\phi(t)$$
with time asymmetric boundary conditions are not allowed, hence we need to modify the mathematical description of the system. 

\section{\label{sec:rhs}Rigged Hilbert Space Topology}
For every fixed $\psi_0\in \Psi$, the translation $T:\Psi\rightarrow\Psi$ such that $\psi\rightarrow \psi+\psi_0$ is a linear homeomorphism of $\Psi$ on itself. Therefore $\tau$ is uniquely determined by the neighborhood system  $I(0)$ centered at the origin, because every other neighborhood of any point $\psi$ of $\Psi$ is obtained by translating a neighborhood of the origin of the vector $\psi$. $(\Psi,\tau)$ is said to be locally convex if $\mathcal{C}=\{C\in I(0)\;\mid\; C \mbox{ is convex}\}$ is a neighborhood local basis. Since every open ball centered at the origin is convex, it is also a member of $\mathcal{C}$ if and only if $\exists A\in\tau\;\mid\;0\in A\subset B_r(0)\;\forall B_r(0)$. By this last condition, we build a locally convex topology $\tau$ on $\Psi$ that is finer than the topology $\tau_d$ induced by the norm.

Let us suppose that $(\Psi,\tau)$ and $(\mathcal{H},\tau_{\mathcal{H}})$ are the previously described spaces and, besides, $\tau$ is locally convex and finer than $\tau_{\mathcal{H}}$. Then we can define another completion $\Phi$ of $\Psi$, this time with respect to $\tau$, and find another complete space $(\Phi,\tau_{\Phi})$ that is different from  $(\mathcal{H},\tau_{\mathcal{H}})$. Precisely, $\Phi\subset\mathcal{H}$, and $\Phi$ is dense in $\mathcal{H}$. Moreover, $\Phi\subset\mathcal{H}\Rightarrow\mathcal{H}^*\subset\Phi^*$, where $\mathcal{H}^*$ and $\Phi^*$ are the dual spaces of $\mathcal{H}$ and $\Phi$, respectively.

The definition of dual space is the basis to build a RHS and we need a more physically accessible dual space, according to \cite{bohm1} and \cite{bohm2}.
Let $\mathcal{E}$ be a Euclidean space. We identify the scalar product on $\mathcal{E}$ as $(\cdot|\cdot)$; instead $\langle\cdot|\cdot\rangle$ is the operatorial product on the dual space $\mathcal{E}^*$, namely $F(v)=\langle F|v\rangle$.
We define our dual space $\Phi^{\times}$ as the space of \textit{antilinear} and continuous functionals on $\Phi$, that is
$$F\in\Phi^{\times} \;\Longleftrightarrow\;F(\phi)=\langle\phi|F\rangle.$$
Thus every functional in $\Phi^{\times}$ has a sort of complex conjugate in $\Phi^*$, and the Riesz-Frechet representation theorem on the Hilbert space $\mathcal{H}$ still works, hence $\mathcal{H}=\mathcal{H}^{\times}$. In this manner we obtain the \textit{Gelfand triplet} $\Phi\subset\mathcal{H}\subset\Phi^{\times}$, which defines our RHS.

\section{\label{sec:eigen} Gamow Vectors}
It is well known that, in order to be observable, the Hamiltonion operator $H$ of a quantum system must be self-adjoint on $\mathcal{H}$, so $H=H^{\dagger}$. Nevertheless $H \neq H^{\times}$ on $\Phi^{\times}$.

Let us consider the secular equation
\begin{equation}
\label{secolare}
H^{\times}|E\rangle = E|E\rangle.
\end{equation}
If $|E\rangle\in\Phi^{\times}\setminus\mathcal{H}$, we cannot affirm that the corresponding eigenvalue $E$ is a real number.
We define a generalized eigenvector $|E\rangle\in\Phi^{\times}$, which has complex eigenvalue, as a \textit{Gamow vector} $|\phi_n^{G}\rangle=|E_n^{\pm}\rangle=|E_R\pm i\frac{\Gamma_n}{2}\rangle$ (subscript $R$ is due to one of the first applications of this theory, that Bohm developed in scattering experiments \cite{bohm1}, and it is related to the resonances of the system). From the Schr\"odinger equation (in units such that $\hbar=1$), we get a unitary operator $U(t)=e^{-iHt}$  for the temporal evolution of any state in $\mathcal{H}$. We see that $U(t)^{\times}=e^{iH^{\times}t}$ is not unitary on $\Phi^{\times}$:
\begin{equation}
\label{evoluzione_gamow}
U(t)^{\times}|E_R\pm i\frac{\Gamma_n}{2}\rangle=e^{i E_R t}e^{\mp \frac{\Gamma_n}{2}t}|E_R\pm i\frac{\Gamma_n}{2}\rangle,
\end{equation}
$U(t)^{\times}$ is not an isometry, because
\begin{equation}
\label{evoluzione_norma}
||U(t)^{\times}|E_R\pm i\frac{\Gamma_n}{2}\rangle||^2=
e^{\mp\Gamma_n t}|||E_R\pm i\frac{\Gamma_n}{2}\rangle||^2.
\end{equation}
Moreover
\begin{equation}
\label{evoluzione_decadimento}
||U(t)^{\times}|E_R\pm i\frac{\Gamma_n}{2}\rangle||\stackrel{t\rightarrow\pm\infty}{\longrightarrow}0
\end{equation}
and
\begin{equation}
\label{evoluzione_esplosione}
||U(t)^{\times}|E_R\pm i\frac{\Gamma_n}{2}\rangle||
\stackrel{t\rightarrow\mp\infty}{\longrightarrow}+\infty.
\end{equation}

In a physical context, we need to identify $\Phi$ with the Schwartz space $S(\mathbb{R}^N)$, that is, the space of rapidly decreasing functions, and the Hilbert space $\mathcal{H}$ with the space of quadratically integrable functions $\mathcal{L}^2(\mathbb{R}^N)$, so these last two expressions suggest that we need to define the following new spaces:
$$
\Phi_-=\left\{\phi\in\Phi\mid f(E)=\langle\phi|E^-\rangle\in S(\mathbb{R})\cap\mathcal{H}_-^2\right\},
$$
$$
\Phi_+=\left\{\phi\in\Phi\mid f(E)=\langle\phi|E^+\rangle\in S(\mathbb{R})\cap\mathcal{H}_+^2\right\};
$$
where $\mathcal{H}_-^2$ and $\mathcal{H}_+^2$ are Hardy spaces bounded from below and from above, respectively.
To sum up, $\Phi_{\pm}$ are dense in $\Phi$, $\Phi=\Phi_-+\Phi_+$ ($\Phi_-\cap\Phi_+\neq0$ generally) and $\Phi$ is dense in $\mathcal{H}$, consequently
\begin{equation}
\label{spazi_meno}
\Phi_-\stackrel{\scriptscriptstyle{dense}}{\subset}\Phi
\stackrel{\scriptscriptstyle{dense}}{\subset}\mathcal{H}
\stackrel{\scriptscriptstyle{dense}}{\subset}\Phi^{\times}
\stackrel{\scriptscriptstyle{dense}}{\subset}
\Phi_-^{\times},
\end{equation}
\begin{equation}
\label{spazi_piu}
\Phi_+\stackrel{\scriptscriptstyle{dense}}{\subset}\Phi
\stackrel{\scriptscriptstyle{dense}}{\subset}\mathcal{H}
\stackrel{\scriptscriptstyle{dense}}{\subset}\Phi^{\times}
\stackrel{\scriptscriptstyle{dense}}{\subset}
\Phi_+^{\times}.
\end{equation}
We have now found two Gelfand triplets, $\Phi_-\subset\mathcal{H}\subset\Phi_-^{\times}$ and $\Phi_+\subset\mathcal{H}\subset\Phi_+^{\times}$, where the evolution operator $U(t)$ acts as a semigroup, because it is well defined and continuous only for $t\leq0$ on $\Phi_-$, and only for $t\geq0$ on $\Phi_+$. The value $t=0$ expresses the intrinsic irreversibility we have when, for example, we divide an experiment into a preparation stage and a registration stage. In this case, $\Phi_-$ will be the space of the initial states and $\Phi_+$ will be the space of the detected states.

\section{\label{sec:damped}Quantization of a Damped Motion}

For its simplicity and its relevance, the \textit{harmonic oscillator} (HO) can be chosen to introduce the study of quantum mechanics in a time symmetric context \cite{chruscinski1,chruscinski2}. The classical HO Hamiltonian is
$$H=\frac{p^2}{2m}+\frac{m\omega^2}{2}x^2.$$
We quantize the HO by converting the canonical coordinates $x,p$ into the operators $\hat{x},\hat{p}$ such that
$$[\hat{x},\hat{p}]=i\hbar,$$
and we find the spectrum of $H$:
$$H\psi(x)=E\psi(x),\;\;E_n=\hbar\omega\left(n+\frac{1}{2}\right),$$
\begin{equation}
\label{HO}
\psi_n(x)=\sqrt[4]{\frac{m\omega}{\hbar\pi}}\frac{1}{\sqrt{2^nn!}}H_n\left(\sqrt{\frac{m\omega}{\hbar}}x\right),
\end{equation}
where $H_n(x)=(-1)^nx^2\frac{d^n}{dx^n}e^{-x^2}$ are the Hermite polynomials.

In a time asymmetric context, considering the equation of a damped motion comes natural for its inherent irreversibility. In fact, if we consider the classical dynamical system in one dimension
\begin{equation}
\label{moto_smorzato}
\left\{\begin{array}{c}\frac{d}{dt}u(t)=-\gamma u(t)\\u(0)=u_0\end{array}\right.
\end{equation}
where $\gamma>0$ and $m=\hbar=1$, we have  
$$u(t)=e^{-\gamma t}u_0,$$
which represents a damping for $t\geq 0$.
We quantize it exactly as we did for the HO, even if this one is not a Hamiltonian system. In a general $n$-dimensional space, one defines a dynamical system as
$$\frac{du}{dt}=X(u),$$
where $X$ is a vector field. Using canonical coordinates $(u^1,...,u^n,v^1,...,v^n)$, we get the Hamiltonian
$$H(u,v)=\sum_{k=1}^n v_kX_k(u),$$
where $X_k$ are the components of $X$ in the coordinate basis, so for Eq.(\ref{moto_smorzato})
$$H(u,v)=-\gamma uv.$$
Since the quantization must take into account that $\hat{v}$ does not commute with $\hat{u}$, we have
\begin{equation}
\label{quantizzazione1}
\hat{H}(\hat{u},\hat{v})=-\frac{\gamma}{2}(\hat{u}\hat{v}+\hat{v}\hat{u}).
\end{equation}
By performing the canonical transformation
\begin{equation}
\label{trasf_can}
\hat{u}=\frac{\gamma\hat{x}-\hat{p}}{\sqrt{2\gamma}},\;\;\;
\hat{v}=\frac{\gamma\hat{x}+\hat{p}}{\sqrt{2\gamma}},
\end{equation}
one obtains the Hamiltonian of the \textit{reversed harmonic oscillator} (RHO):
\begin{equation}
\label{quantizzazione2}
\hat{H}(\hat{x},\hat{p})=\frac{\hat{p}^2}{2}-\frac{\gamma^2\hat{x}^2}{2}.
\end{equation}
Let us compare the HO and the RHO. We pass from the first one to the second one, by changing $\omega$ into the complex value $i\gamma$ \cite{barton}. This simple transformation allows to move from a parabolic potential bounded from below to a parabolic barrier. This potential overturning produces a completely different physics: the HO models the behavior of a pointlike mass around a stable equilibrium and the RHO gives the dynamics around an unstable equilibrium, an intrinsically irreversible evolution (at variance with an oscillator, a falling body never goes back to its initial position).

In this section, we analyze the Hamiltonian of the damped motion, defined in (\ref{quantizzazione1}). 
As proved in \cite{chruscinski1}, $\hat{H}(\hat{u},\hat{v})$ is self-adjoint on $\mathcal{L}^2(\mathbb{R})$ and parity invariant. We define the time reversal operator $T$ such that
$$T\phi(t):=\phi(-t)\;\;\Rightarrow TU(t)=U^{\dagger}(t)T\;\;\Rightarrow U(t)TU(t)=T,$$
where $U(t):=e^{-iHt}$. $T$ plays a fundamental role in this system, and coincides with the inverse Fourier transformation, i.e. $T\phi(u,t):=\check{F}[\phi](u,t)$,where
$$\check{F}[\phi](x,t)=\frac{1}{\sqrt{2\pi}}\int_{\mathbb{R}}e^{ikx}\phi(k,t)dk.$$

Let us define two families of tempered distributions in $\Phi^{\times}$, the first one

$$\hat{u}|f_0^-\rangle:=0,\;\;\;\;f_0^-(u)=\delta(u),$$

\begin{equation}
\label{fnmeno}
\forall n\in\mathbb{N}\;\;\;
|f_n^-\rangle:=\frac{(-i)^n}{\sqrt{n!}}\hat{v}^n|f_0^-\rangle\;\;\Rightarrow
\end{equation}
$$\Rightarrow\;\; f_n^-(u)=\frac{(-1)^n}{\sqrt{n!}}\frac{d^n}{du^n}\delta(u);$$
and the second one

$$\hat{v}|f_0^+\rangle:=0,\;\;\;\;f_0^+(u)=1,$$

\begin{equation}
\label{fnpiu}
\forall n\in\mathbb{N}\;\;\;
|f_n^+\rangle:=\frac{1}{\sqrt{n!}}\hat{u}^n|f_0^+\rangle\;\;\Rightarrow\;\;
f_n^+(u)=\frac{u^n}{\sqrt{n!}}.
\end{equation}
Hereafter, following \cite{bohm1,bohm2,bohm3}, we denote a tempered distribution $f_n^{\pm}$ a \textit{resonance}. We can see that
$$H^{\times}|f_n^{\pm}\rangle=\pm E_n |f_n^{\pm}\rangle,$$
where $E_n:=i\gamma\left(n+\frac{1}{2}\right)\in\mathbb{C}$.
Given that $f_n^{\pm}$ are tempered distributions, their inverse Fourier transforms are well defined, and they are 

\begin{equation}
\label{trasformatefnmeno}
\check{F}[f_n^-]=\frac{i^n}{\sqrt{2\pi}}f_n^+,
\end{equation}

\begin{equation}
\label{trasformatefnpiu}
\check{F}[f_n^+]=i^n\sqrt{2\pi}f_n^-.
\end{equation}
We show  the \textit{quasi-orthogonality} and the \textit{quasi-completeness} of the resonances:

$$\langle f_n^-|f_m^+\rangle=\delta_{n,m},$$
$$\sum_{n=0}^{\infty}f_n^-(u)f_n^+(y)=\delta(u-y).$$

In order to find real energy values, we need to analyze also the continuous spectrum.
Since $H$ is parity invariant, each generalized eigenvalue is doubly degenerate, thus
$$H^{\times}\psi_{\pm}^E=E\psi_{\pm}^E.$$
As one can see in \cite{chruscinski1}, the generalized eigenfunctions are
\begin{equation}
\label{autofunzioni1}
\psi_{\pm}^E(u)=\frac{1}{\sqrt{2\pi\gamma}}u_{\pm}^
{-\left(\frac{iE}{\gamma}+\frac{1}{2}\right)},
\end{equation}
where $u_{\pm}^{\lambda}$ are tempered distributions such that

$$u_+^{\lambda}:=\left\{\begin{array}{cc}u^{\lambda} & u\geq0\\0 & u<0\end{array}\right.,$$

$$u_-^{\lambda}:=\left\{\begin{array}{cc}0 & u<0\\u^{\lambda} & u\leq0\end{array}\right..$$
It is possible to prove both the orthonormality and the completeness of the eigenfunctions, namely

$$\sum_{\pm}\int[\psi_{\pm}^{E_1}(u)]^*\psi_{\pm}^{E_2}(u)du=\delta(E_1-E_2);$$

$$\sum_{\pm}\int[\psi_{\pm}^{E}(u)]^*\psi_{\pm}^{E}(u')dE=\delta(u-u').$$
Therefore we can apply the Gelfand-Maurin theorem \cite{gadella} and write any function in $S(\mathbb{R})$ as

$$\phi(u)=\sum_{\pm}\int\psi_{\pm}^{E}(u)\langle\phi|\psi_{\pm}^{E}\rangle^* dE.$$

By repeating the same reasoning
\begin{equation}
\label{autofunzioni2}
H^{\times}\check{F}\left[\psi_{\pm}^{-E}\right]=E\check{F}\left[\psi_{\pm}^{-E}\right],
\end{equation}
so one can prove also the orthonormality and the completeness of the inverse Fourier transforms of the eigenfunctions, whence
\begin{equation}
\label{gelfand_fourier}
\phi(u)=\sum_{\pm}\int\check{F}\left[\psi_{\pm}^{-E}\right](u)\langle\phi|\check{F}\left[\psi_{\pm}^{-E}\right]\rangle^* dE.
\end{equation}

We have just defined two groups of eigenfunctions, $\psi_{\pm}^E(u)$ and $\check{F}\left[\psi_{\pm}^{-E}\right](u)$, which represent the continuous spectrum of the Hamiltonian of a damped motion into the RHS. Moreover, we have just seen that they depend on the tempered distributions $u_{\pm}^{-\left(\frac{iE}{\gamma}+\frac{1}{2}\right)}$, which have simple poles in the complex plane when
$$E=-E_n=-i\gamma\left(n+\frac{1}{2}\right).$$
Thanks to the properties of the generalized function $u_{\pm}^{\lambda}$ \cite{chruscinski1}, we can finally state what follows:
\begin{equation}
\label{res_psi}
Res\left[\psi_{\pm}^{E},-E_n\right]=\frac{(\pm 1)^n i \sqrt{\gamma}}{\sqrt{2\pi n!}}f_n^-,
\end{equation}
\begin{equation}
\label{res_F}
Res\left[\check{F}[\psi_{\pm}^{-E}],E_n\right]=\frac{(\pm i)^n i \sqrt{\gamma}}{2\pi\sqrt{n!}}f_n^+.
\end{equation}
By defining the following spaces, we get two Gelfand triplets:
$$\mathcal{H}=\mathcal{L}^2(\mathbb{R}),$$
$$\Phi=S(\mathbb{R}),$$
\begin{equation}
\label{phimeno}
\Phi_-=\left\{\phi\in\Phi\mid f(E)=\langle\phi|\check{F}[\psi_{\pm}^{-E}]\rangle\in \mathcal{H}_-^2\right\},
\end{equation}
\begin{equation}
\label{phipiu}
\Phi_+=\left\{\phi\in\Phi\mid f(E)=\langle\phi|\psi_{\pm}^E\rangle\in \mathcal{H}_+^2\right\},
\end{equation}
where $\mathcal{H}_+^2$ ($\mathcal{H}_+^2$) is the Hardy space on the upper (lower) complex half-plane, respectively.

From this framework into the RHS $\Phi^{\times}$, we can infer the irreversible evolution of certain waves in $\Phi$.
We established above the connection between the continuous and the point spectrum. Now we make this link definitively clear and we show that the evolution operator acts as a semigroup on $\Phi_{\pm}$ for a well-defined orientation of the arrow of time. By recalling Eqs.(\ref{res_psi}) and (\ref{res_F}), we apply the residue theorem to initial data in $\Phi_{\pm}$ \cite{chruscinski1} and get two different expansions in GV:

$$\phi^+(u)=\sum_{n=0}^{+\infty}\langle\phi^+|f_n^+\rangle f_n^-(u)\;\;\forall\;\phi^+\in \Phi_+;$$
$$\phi^-(u)=\sum_{n=0}^{+\infty}\langle\phi^-|f_n^-\rangle f_n^+(u)\;\;\forall\;\phi^-\in \Phi_-.$$

Thanks to the following definitions of two new function spaces, both of them subspaces of $S(\mathbb{R})$ and isomorphic by the inverse Fourier transformation, we can establish the relation between $\Phi_+$ and $\Phi_-$:\newline
$\mathcal{D}=C_c^{\infty}(\mathbb{R})$ is the space of the infinitely differentiable functions with compact support; \newline
$\mathcal{Z}=\left\{\check{F}[\phi]\,|\,\phi\in\mathcal{D}\right\}$, where $\check{F}$ is the inverse Fourier transformation.\newline
Since for each function $\phi\in\mathcal{Z}$, we have
$$\phi(u)=\sum_{n=0}^{+\infty}\frac{1}{n!}\frac{d^n}{du^n}\phi(u)|_{u=0}u^n=\sum_{n=0}^{+\infty}f_n^+(u)\langle f_n^-|\phi\rangle,$$
while, at the same time, every $\psi\in\mathcal{D}$ is the Fourier transform of a function in $\mathcal{Z}$, hence
$$\psi(u)=\frac{1}{\sqrt{2\pi}}\int_{\mathbb{R}}\check{F}[\psi](v)e^{-ivu}dv=\sum_{n=0}^{+\infty}f_n^-(u)\langle f_n^+|\psi\rangle.$$
We can state that
\begin{equation}
\label{equivalenze}
\Phi_+\equiv\mathcal{D},\;\;\Phi_-\equiv\mathcal{Z}.
\end{equation}

At last, we study the evolution operator $U(t)=e^{-i Ht}$.
$U$ is a unitary group on $\mathcal{H}=\mathcal{L}^2(\mathbb{R})$,  given that if $\psi(u,0)\in\mathcal{L}^2(\mathbb{R})$ then
\begin{equation}
\label{damped_evolution}
\psi(u,t)=U(t)\psi(u,0)=e^{\frac{\gamma}{2}t}\psi(e^{\gamma t}u,0),
\end{equation}
transformation that turns out to be an isometry on $\mathcal{L}^2(\mathbb{R})$. This means that if $\psi(u,t)$ solves the Schr\"odinger equation, then also $T\psi(u,t)=\psi(u,-t)$ does.
Therefore the theory is time-reversal invariant on the Hilbert space $\mathcal{H}$, without letting us see the damping we expected. Where do we observe the temporal irreversibility?
It lacks the analysis of $U$ restricted to $\Phi_{\pm}$. If $\phi^+(u,0)\in\Phi_+$ then
$$\langle U(t)\phi^+|\psi_{\pm}^E\rangle=\langle\phi^+|U^{\times}(t)\psi_{\pm}^E\rangle=$$
$$=e^{iEt}\langle\phi^+|\psi_{\pm}^E\rangle\;\in\mathcal{H}^2_+\;\;\Leftrightarrow\;\;t\geq 0;$$
on the other hand,  if $\phi^-(u,0)\in\Phi_-$ then
$$\langle U(t)\phi^-|\check{F}[\psi_{\pm}^{-E}]\rangle=\langle U(-t)\check{F}[\phi^-]|\psi_{\pm}^{-E}\rangle=$$
$$=\langle\check{F}[\phi^+]|U^{\times}(-t)\psi_{\pm}^{-E}\rangle=e^{iEt}\langle\check{F}[\phi^-]|\psi_{\pm}^{-E}\rangle=$$
$$=e^{iEt}\langle\phi^-|\check{F}[\psi_{\pm}^{-E}]\rangle\;\in\mathcal{H}^2_-\;\;\Leftrightarrow\;\;t\leq 0.$$
We conclude that $U(t)$ establishes two semigroups:
$$U_+(t):\Phi_+\longrightarrow\Phi_+\;\;\;\forall t\geq 0$$
and
$$U_-(t):\Phi_-\longrightarrow\Phi_-\;\;\;\forall t\leq 0.$$
We have just found a way to model irreversible phenomena. In fact, the action of $U$ allows to choose an orientation of the temporal arrow: if it goes forward from zero, then our initial data is in $\Phi_+$, otherwise it is in $\Phi_-$, indeed
$$\phi^+(u,t)=\sum_ne^{-\gamma(n+1/2)t}\langle\phi^+|f_n^+\rangle f_n^-(u)$$
and
$$\phi^-(u,t)=\sum_ne^{\gamma(n+1/2)t}\langle\phi^-|f_n^-\rangle f_n^+(u).$$
Moreover, all the physics we get fixing a specific orientation of time's arrow is achievable fixing the other one too, because time reversal operator $T$ establishes an isomorphism between $\Phi_+$ and $\Phi_-$, in fact
$$T\phi^+(u,t)=U(-t)T\phi^+(u,0)=\phi^-(u,-t).$$

Summarizing, we got an irreversible quantum system by observing that the evolution operator acts as a semigroup on $\Phi_{\pm}$, due to the presence of resonant states $f_n^{\pm}$. In this way, the instant $t=0$ separates the evolution in two complementary directions: if one starts from $\Phi_+$, one can stays forever in $\Phi_+$ only evolving forward in time. In other words one chooses the temporal orientation, fixes the signature of $\Phi_{\pm}$, and cannot go backwards.

\begin{figure}[ht!]
\label{tripletta}
\begin{center}
\includegraphics[width=0.5\textwidth]{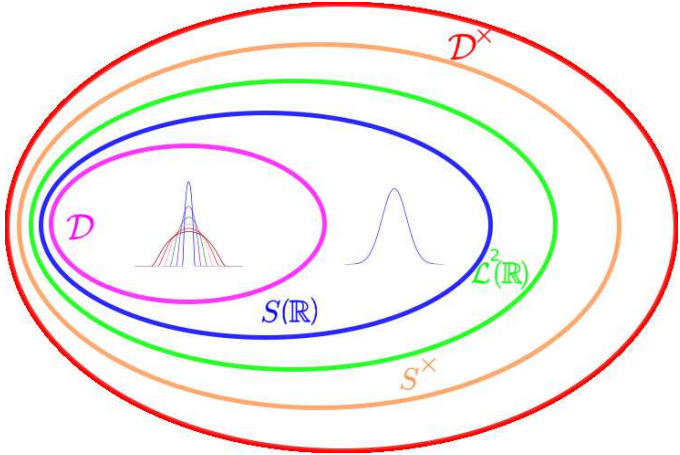}
\caption{(Color online) Pictorial representation of Gelfand triplet defined in Eq.(\ref{spazi_piu}). Here $\Phi_+\equiv\mathcal{D}$, $\Phi\equiv S(\mathbb{R})$ and $\mathcal{H}\equiv\mathcal{L}^2(\mathbb{R})$. One can get an Euler-Venn diagram also for the triplet in Eq.(\ref{spazi_meno}) by changing $\Phi_+$ with $\Phi_-$ and $\mathcal{D}$ with $\mathcal{Z}$.}
\end{center}
\end{figure}

\section{\label{sec:rho}The Reversed Harmonic Oscillator:\protect\\Remarkable Results}

We consider the family of operators \cite{chruscinski2}
$$\hat{V}_{\lambda}=\exp\left\{\frac{\lambda}{2}(\hat{x}\hat{p}+\hat{p}\hat{x})\right\}.$$
In a system of measurement where $\hbar=1$, we have \newline $[\hat{x},\hat{p}]=i$, so
$$\hat{V}_{\lambda}\phi(x)=e^{-i\frac{\lambda}{2}}\phi(e^{-i\lambda}x),$$
whence
$$\hat{V}_{\lambda}\hat{x}\hat{V}_{\lambda}^{-1}=e^{-i\lambda}\hat{x}$$
and
$$\hat{V}_{\lambda}\hat{p}\hat{V}_{\lambda}^{-1}=e^{i\lambda}\hat{p}.$$
If we recall the Hamiltonian in the equation (\ref{HO}) 
$$H_{HO}:=\frac{\hat{p}^2}{2}+\frac{\gamma^2\hat{x}^2}{2},$$
it is easy to see that
$$\hat{V}_{\pm\frac{\pi}{4}} H \hat{V}_{\pm\frac{\pi}{4}}^{-1}=\pm i H_{HO}$$
and we can transform the results we already know for the HO in results for the RHO:
$$E_n^{HO}=\gamma(n+\frac{1}{2}),\,  E_n=i E_n^{HO}\,\in\,\mathbb{C}$$
$$\psi_n^{HO}=\left(\frac{\gamma}{\pi}\right)^{1/4}(2^n n!)^{-1/2}e^{-\frac{\gamma}{2}x^2}H_n(\sqrt{\gamma}x)$$
$$f_n^{\pm}=\hat{V}_{\pm\frac{\pi}{4}}^{-1}\psi_n^{HO}\in S^{\times}(\mathbb{R}).$$

\subsection{\label{subs:unitary}The Unitary Transformation:\protect\\from $(u,v)$ to $(x,p)$ Framework}

One passes from the HO to the RHO through the operator $\hat{V}_{\pm\frac{\pi}{4}}$, but can also pass from $H(\hat{u},\hat{v})$ to $H(\hat{x},\hat{p})$, i.e. from the damped motion to the RHO, through a canonical transformation and find a relation between the spectra of these two Hamiltonians. 

The canonical transformation from $(u,v)$ to $(x,p)$ it is generated by the generating function
\begin{equation}
\label{generatrice}
S(x,u)=\frac{\gamma}{2}x^2-\sqrt{2\gamma}xu+\frac{1}{2}u^2,
\end{equation}
with $p=\frac{\partial S}{\partial x}$, $v=-\frac{\partial S}{\partial u}$.

We define the unitary transformation
\begin{equation}
  \label{trasformazione}
  \mathcal{U}:\mathcal{L}^2(\mathbb{R})\longrightarrow\mathcal{L}^2(\mathbb{R})
\end{equation}
such that
$$f(u)\longrightarrow(\mathcal{U}f)(x)=\tilde{C}\int_{\mathbb{R}}f(u)e^{i S(x,u)}du,$$
with $\tilde{C}:=e^{-i\frac{\pi}{8}}\sqrt[4]{\frac{\gamma}{2\pi^2}}$ and we can prove that $\mathcal{U}$ is unitary by demonstrating that
$$|\tilde{C}|^2\int_{\mathbb{R}}e^{i [S(x,u)-S(x',u)]}du=\delta(x-x').$$

In order to get a relation of quasi-orthogonality and quasi-completeness for the resonances, we need to understand the nature of the operator $\hat{V}_{\lambda}$. It acts almost like the evolution operator $U$ in Eq.(\ref{damped_evolution}), with a complex (instead of real) exponential, but this is enough only to say that $\hat{V}_{\lambda}$ is unitary for pure imaginary $\lambda$, not for every $\lambda\in\mathbb{C}$. In fact, for a generic $\lambda=\omega+i\gamma$, where $\omega,\gamma\in\mathbb{R}$, one has
$$\langle\hat{V}_{\lambda}\phi|\hat{V}_{\lambda}\psi\rangle=\int_{\mathbb{R}}dx\left[e^{\frac{\gamma-i\omega}{2}}\phi\left(e^{\gamma-i\omega}x\right)\right]^*e^{\frac{\gamma-i\omega}{2}}\psi\left(e^{\gamma-i\omega}x\right)=$$
$$=e^{i\omega}\int_{\mathbb{R}}dx\left[\phi(x)\right]^*\psi(x)=e^{i\omega}\langle\phi|\psi\rangle.$$
Therefore it is not surprising that $f_n^{\pm}$ are only proportional to $\mathcal{U}[f_n^{\pm}(u)](x)$ and not exactly equal. In fact
$$f_n^{\pm}(x)=e^{in\frac{\pi}{4}}(2\pi)^{\pm\frac{1}{4}}\mathcal{U}[f_n^{\pm}(u)](x).$$ 
Nevertheless, we achieve the same relation of quasi-orthogonality and quasi-completeness we had before:
$$\langle f_n^{\pm}(x)| f_m^{\mp}(x) \rangle =\delta_{nm};$$
$$\sum_{n=0}^{+\infty}\left[f_n^{\pm}(x)\right]^*f_n^{\mp}(x')=\delta(x-x').$$
Moreover
$$\left[f_n^{\pm}(x)\right]^*=f_n^{\mp}(x).$$
Recalling the equations (\ref{trasformatefnmeno}), (\ref{trasformatefnpiu}) and the meaning of the inverse Fourier transform for the damped motion represented by $\hat{H}(\hat{u},\hat{v})$ (the inverse Fourier transform coincides with the time reversal operator $T$ in that system), one has $T=C$, where $C$ is the complex conjugation operator, as shown in \cite{chruscinski2}.

We want to find $\chi^E$ such that
$$H\chi^E=E\chi^E.$$
From \cite{chruscinski2} we get the complete derivation of the following solutions:

$$\chi_+^E(x)=\frac{\tilde{C}}{\sqrt{2\pi\gamma}}i^{\frac{\nu+1}{2}}\Gamma(\nu +1)D_{-\nu-1}(-\sqrt{-2\gamma i}x),$$
$$\chi_-^E(x)=\chi_+^E(-x),$$
where here $\nu=-\left(i\frac{E}{\gamma}+\frac{1}{2}\right)$ and
$$D_{\nu}(z):=\frac{e^{-\frac{z^2}{4}}}{\Gamma(-\nu)}\int_{\mathcal{R}}\xi_{\pm}^{-\nu-1}e^{\mp z\xi-\frac{1}{2}\xi^2}d\xi$$
is a Whittaker function \cite{gradshteyn}.

If one remembers the equation (\ref{autofunzioni2}), one knows that the set of eigenfunction is not complete yet. In fact, the two families of functions $\eta_{\pm}^E(x):=\left(\mathcal{U}\check{F}[\psi_{\pm}^{-E}]\right)(x)$ still miss, and we obtain
$$H\eta_{\pm}^E=-E\eta_{\pm}^E,$$
$$\eta_+^E(x)=\frac{\tilde{C}}{\sqrt{2\pi\gamma}}i^{\frac{\nu+1}{2}}\Gamma(-\nu)D_{\nu}(-\sqrt{2\gamma i}x),$$
$$\eta_-^E(x)=\eta_+^E(-x).$$
We observe that
$$\eta_{\pm}^E(x)=[\chi_{\pm}^E(x)]^*,$$
fact which confirms that the time reversal operator $T$ acts like the complex conjugation $C$.

From the corresponding properties satisfied by $\psi_{\pm}^E(u)$ and from the unitary nature of $\mathcal{U}$ we have
$$\sum_{\pm}\int_{\mathbb{R}}[\chi_{\pm}^E(x)]^*\chi_{\pm}^{E'}(x)dx=\delta(E-E');$$
$$\sum_{\pm}\int_{\mathbb{R}}[\chi_{\pm}^E(x)]^*\chi_{\pm}^{E}(x')dE=\delta(x-x');$$
$$\sum_{\pm}\int_{\mathbb{R}}[\eta_{\pm}^E(x)]^*\eta_{\pm}^{E'}(x)dx=\delta(E-E');$$
$$\sum_{\pm}\int_{\mathbb{R}}[\eta_{\pm}^E(x)]^*\eta_{\pm}^{E}(x')dE=\delta(x-x').$$

At this point, we have all the tools we need to study the analytic properties of these four families of eigenfunctions. The outcome is that $\chi_{\pm}^E(x)$ and $\eta_{\pm}^E(x)$ have simple poles at $E=-E_n$ and $E=E_n$, respectively. Furthermore,
$$Res[\chi_{\pm}^E(x);-E_n]=\frac{\tilde{C}}{\sqrt{2\pi\gamma}}\frac{(-1)^n}{n!}i^{-\frac{n}{2}}D_n(\mp\sqrt{-2\gamma i}x),$$
$$Res[\eta_{\pm}^E(x);E_n]=\frac{\tilde{C}}{\sqrt{2\pi\gamma}}\frac{(-1)^n}{n!}i^{\frac{n+1}{2}}D_n(\mp\sqrt{2\gamma i}x).$$

In \cite{gradshteyn}, \cite{morse} and \cite{abramowitz} one can find out that \newline $D_n(y)=2^{-\frac{n}{2}}e^{-\frac{y^2}{4}}H_n\left(\frac{z}{\sqrt{2}}\right)$. This, together with $H_n(-y)=(-1)^nH_n(y)$, allows us to obtain
$$Res[\chi_{\pm}^E(x);-E_n]\propto f_n^+(x)$$
and
$$Res[\eta_{\pm}^E(x);E_n]\propto f_n^-(x).$$

Following section \ref{sec:damped}, we get $\Phi_{\pm}$ from the residues of the RHO eigenfunctions:
$$\mathcal{H}=\mathcal{L}^2(\mathbb{R}),$$
$$\Phi=S(\mathbb{R}),$$
$$\Phi_-=\left\{\phi\in\Phi\mid f(E)=\langle\phi|\eta_{\pm}^E\rangle\in \mathcal{H}_-^2\right\},$$
$$\Phi_+=\left\{\phi\in\Phi\mid f(E)=\langle\phi|\chi_{\pm}^E\rangle\in \mathcal{H}_+^2\right\}.$$

\subsection{\label{subs:semigroup} The Evolution Operator Acting like a Semigroup}

We study waves $\phi_{\pm}\in\Phi_{\pm}$ and the action of the evolution operator. We have $T=C$ and
$$T(\Phi_+)=\Phi_-.$$

Writing envelopes of $\phi_{\pm}$ in series of resonances:
$$\phi^+(x)=\sum_{n=0}^{+\infty}\langle\phi^+|f_n^+\rangle^* f_n^-(x)\;\;\forall\;\phi^+\in \Phi_+;$$
$$\phi^-(x)=\sum_{n=0}^{+\infty}\langle\phi^-|f_n^-\rangle^* f_n^+(x)\;\;\forall\;\phi^-\in \Phi_-.$$
Thanks to the Gelfand-Maurin spectral theorem
$$\phi^+(x)=\sum_{\pm}\int_{\mathbb{R}}dE\chi_{\pm}^E(x)\langle\phi^+|\psi_{\pm}^E\rangle^*$$
and
$$\phi^-(x)=\sum_{\pm}\int_{\mathbb{R}}dE\eta_{\pm}^E(x)\langle\phi^-|\eta_{\pm}^E\rangle^*.$$

In conclusion, even in this case, the temporal evolution operator $U(t)=e^{-i Ht}$ establishes a unitary group on $\mathcal{H}=\mathcal{L}^2(\mathbb{R})$, and two semigroups:
$$U_+(t):\Phi_+\longrightarrow\Phi_+\;\;\;\forall t\geq 0;$$
$$U_-(t):\Phi_-\longrightarrow\Phi_-\;\;\;\forall t\leq 0.$$

Furthermore, if $\phi^+(x,0)\in\Phi_+$ then
$$\phi^+(x,t)=\sum_ne^{-\gamma(n+1/2)t}\langle\phi^+|f_n^+\rangle^* f_n^-(x),$$
while,  if $\phi^-(x,0)\in\Phi_-$ then
$$\phi^-(x,t)=\sum_ne^{\gamma(n+1/2)t}\langle\phi^-|f_n^-\rangle^* f_n^+(x).$$

We stress again that we got an irreversible quantum theory by studying the action of $U$ on $\Phi_{\pm}$ as a semigroup. Time $t=0$ splits the evolution in two diametrically opposed directions, and it becomes the instant which separates two different dynamics.

\section{\label{sec:compact}Functions with Compact Support}

In this section we examine a function set in $\Phi_+$. We start working in the $(u,v)$ representation, where $H(\hat{u},\hat{v})=-\frac{\gamma}{2}[\hat{u}\hat{v}+\hat{v}\hat{u}]$. We analyse the evolution in $(u,v)$ and $(x,p)$ planes. In $(x,p)$ the Hamiltonian is $H=\frac{\hat{p}^2}{2}-\frac{\gamma^2\hat{x}^2}{2}$ (we fix $\gamma=1$ hereafter).

\subsection{\label{subs:uv} Wave Packets in $(u,v)$ Plane}
 We previously proved that $\Phi_+$ and $\Phi_-$ coincide with $\mathcal{D}$ and $\mathcal{Z}$, respectively. We choose the forward orientation of the temporal arrow, so we focus our attention on the triplet
$$\Phi_+\subset\mathcal{H}\subset\Phi_+^{\times},$$
that is $\mathcal{D}\subset\mathcal{L}^2(\mathbb{R})\subset\mathcal{D}^{\times}$.

Let us consider the family of functions
\begin{equation}
\label{campane}
\phi_{\epsilon}(u)=\left\{\begin{array}{cr}K_{\epsilon}\exp\left[\frac{1}{\left(\frac{u}{\epsilon}\right)^2-1}\right]&|u|<\epsilon\\0&|u|\geq\epsilon\end{array}\right.
\end{equation}
where $\epsilon>0$ and $K_{\epsilon}$ is such that $||\phi_{\epsilon}||_2=1$, i.e. $\left(\int_{\mathbb{R}}|\phi_{\epsilon}(u)|^2dx\right)^{\frac{1}{2}}=1$. $\phi_{\epsilon}(u)$ is a function of class $C^{\infty}(\mathbb{R})$, precisely
$$\phi_{\epsilon}(u)\in\mathcal{D}\;\;\forall\,\epsilon>0.$$

\begin{figure}[ht!]
\label{campanegraph}
\begin{center}
\includegraphics[width=0.45\textwidth]{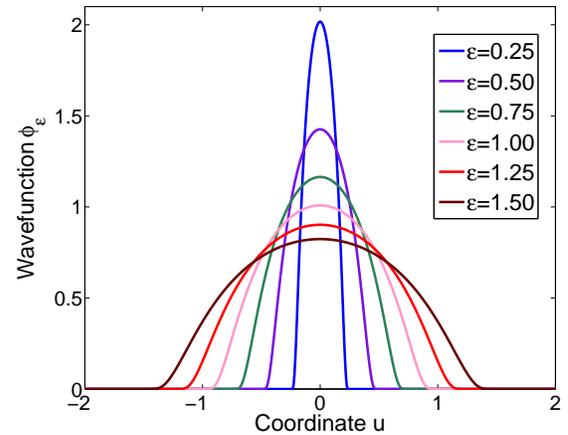}
\caption{(Color online) Functions $\phi_{\epsilon}$ with compact support defined in Eq.(\ref{campane}) for several $\epsilon$ values.}
\end{center}
\end{figure}

Starting from
$$\sum_{n=0}^{\infty}f_n^-(u)f_n^+(w)=\delta(u-w),$$
with $f_n^-(u)=\frac{(-1)^n}{\sqrt{n!}}\frac{d^n}{du^n}\delta(u)$ and  $f_n^+(u)=\frac{u^n}{\sqrt{n!}}$, we have
$$\phi_{\epsilon}(u)=\int_{\mathbb{R}}dw\delta(u-w)\phi_{\epsilon}(w)=$$
\begin{equation}
\label{swap}
=\int_{\mathbb{R}}dw\sum_{n=0}^{\infty}f_n^-(u)f_n^+(w)\phi_{\epsilon}(w)=\sum_{n=0}^{\infty}f_n^-(u)\langle\phi_{\epsilon}|f_n^+\rangle,
\end{equation}
since $\phi_{\epsilon}\in\mathcal{D}$. In deriving Eq.(\ref{swap}), as discussed in sec.\ref{sec:damped} and in \cite{chruscinski2}, the residue theorem allows to swap the integral and the summation. This is not valid for general functions in $\Phi$ not belonging to $\Phi_+$. We define the N-order \textit{background function} as 
$$\phi_N^{BG}(u,t):=\phi(u,t)-\sum_{n=0}^{N}f_n^-(u)\langle U(t)\phi|f_n^+\rangle^*\;\in\;\Phi^{\times};$$
consequently
$$\phi(u,t)=\sum_{n=0}^{N}f_n^-(u)\langle U(t)\phi|f_n^+\rangle^*+\phi_N^{BG}(u,t)\;\;\; \forall\phi\in\Phi.$$

For $\phi\in\Phi_+$, $\phi_{N\rightarrow +\infty}^{BG}=0$ and $U(t)$ acts as a semigroup. The evolution is a superposition of exponentially decaying functions.
On the contrary, for $\phi\notin\Phi_+$, $\phi_{N\rightarrow +\infty}^{BG}$ does not converge and the evolution includes non exponentially decaying components.

\begin{figure}[ht!]
\begin{center}
\includegraphics[width=0.45\textwidth]{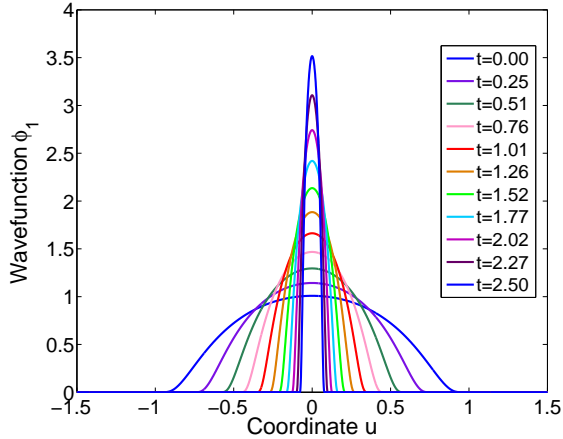}
\caption{(Color online)  One-dimensional evolution of $|\phi_1(u,t)|$ [Eq.(\ref{campane}) with $\epsilon=1$].}
\label{campanauv}
\end{center}
\end{figure}

\begin{figure}[ht!]
\begin{center}
\includegraphics[width=0.45\textwidth]{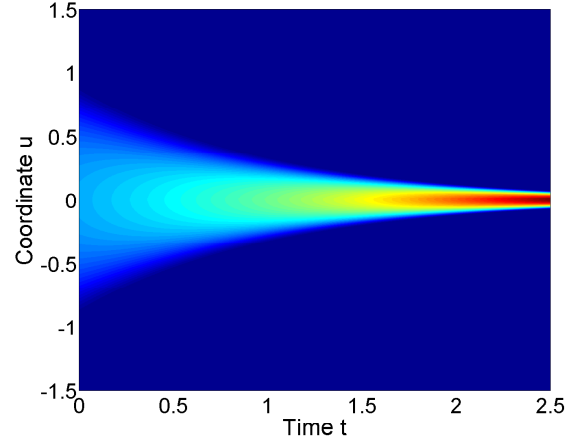}
\caption{(Color online)  Evolution of $|\phi_1(u,t)|$ [Eq.(\ref{campane}) with $\epsilon=1$].}
\label{campanauv2}
\end{center}
\end{figure}

We numerically simulate the Schr\"odinger equation $i\frac{\partial \psi}{\partial t}=H\psi$ for the Hamiltonian $H=i\gamma\left(u\frac{\partial}{\partial u}+\frac{1}{2}\right)$ (with $\gamma$=1), with initial condition $\psi(u,t=0)=\phi_{\epsilon}(u)$. Figures \ref{campanauv} and \ref{campanauv2} show the resulting ``focusing'' evolution.

\begin{figure}[ht!]
\begin{center}
\includegraphics[width=0.45\textwidth]{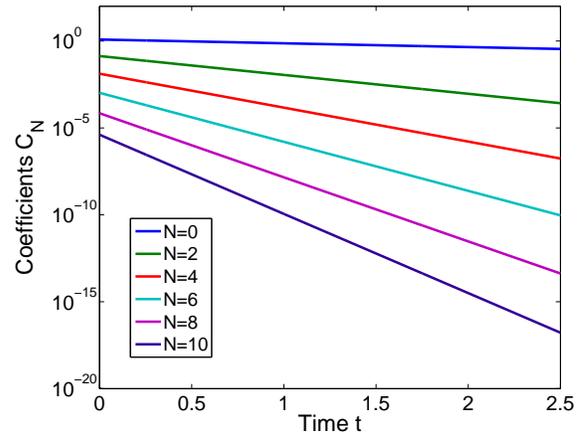}
\caption{(Color online)  Numerically calculated projections $C_N(t):=\langle U(t)\phi_1|f_N^+\rangle^*$ on the N order resonances of a function with compact support [Eq.(\ref{campane}) with $\epsilon=1$] in the (u,v) representation, in a semilogarithmic scale.}
\label{campanacoeffuv}
\end{center}
\end{figure}

Figure \ref{campanacoeffuv} reports the evolution of the coefficients  $C_N(t):=\langle U(t)\phi_1|f_N^+\rangle^*$. These brackets exponentially decay, with quantized decay rates. Into a semilogarithmic scale, the decay rates correspond to straight lines with different slopes. 

\subsection{\label{subs:xp} Wave Packets in $(x,p)$ Plane}
We pass from the $(u,v)$ to $(x,p)$ by the unitary transformation $\mathcal{U}$:
\begin{figure}[ht!]
\begin{center}
\includegraphics[width=0.45\textwidth]{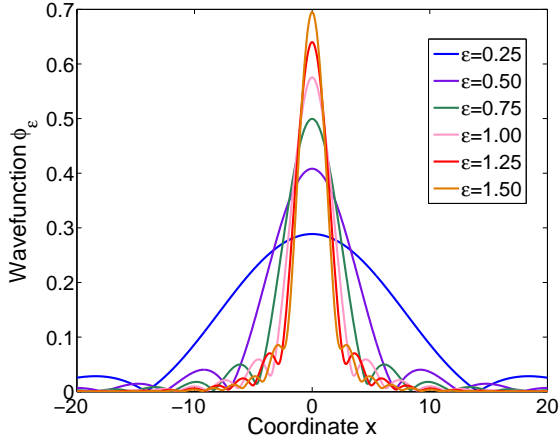}
\caption{(Color online)  Transformed $\phi_{\epsilon}$ [Eq.(\ref{campanatrans}) for various $\epsilon$].}
\label{campanazerostrana}
\end{center}
\end{figure}
$$\phi_{\epsilon}(x)=\mathcal{U}[\phi_{\epsilon}(u)](x)=\sum_{n=0}^{\infty}\mathcal{U}[f_n^-(u)](x)\langle\mathcal{U}\phi_{\epsilon}|\mathcal{U}f_n^+\rangle=$$
\begin{equation}
\label{campanatrans}
=\sum_{n=0}^{\infty}f_n^-(x)\langle\phi_{\epsilon}|f_n^+\rangle,
\end{equation}
with $f_n^{\pm}(x)=\hat{V}_{\pm\frac{\pi}{4}}^{-1}\psi_n^{HO}(x)$.

We numerically analyse the transformed functions. In Fig. \ref{campanazerostrana}, one can see several $\left(\mathcal{U}\phi_{\epsilon}\right)(x)$. We remark that functions $\phi_{\epsilon}$, which have compact support in $(u,v)$, do not have compact support in $(x,p)$ phase plane.  
\begin{figure}[ht!]
\begin{center}
\includegraphics[width=0.45\textwidth]{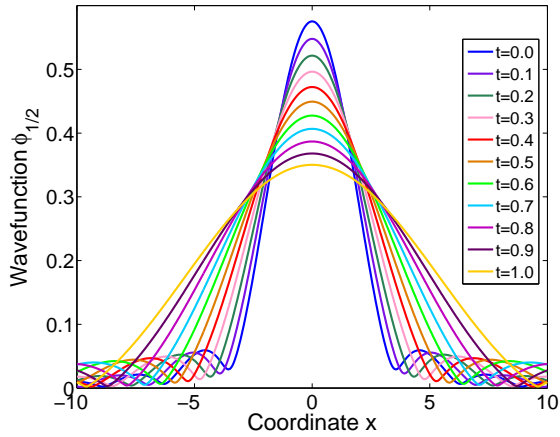}
\caption{(Color online)  One-dimensional evolution of a transformed function with compact support [Eq.(\ref{campanatrans}), $\epsilon=1/2$] with a RHO potential.}
\label{campanastrana}
\end{center}
\end{figure}

\begin{figure}[ht!]
\begin{center}
\includegraphics[width=0.45\textwidth]{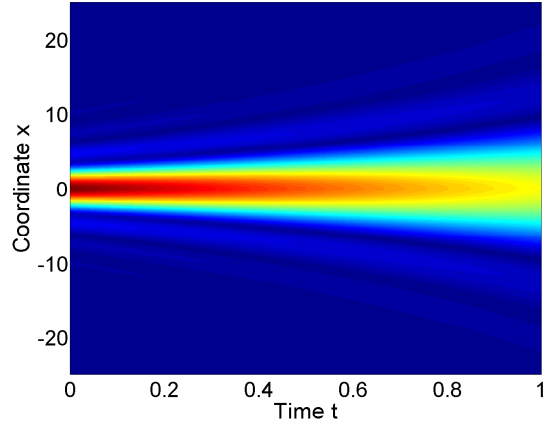}
\caption{(Color online)  Evolution of a transformed function with compact support [Eq.(\ref{campanatrans}), $\epsilon=1/2$] with a RHO potential.}
\label{campanastrana2}
\end{center}
\end{figure}

We numerically study the evolution of wave packets in $(x,p)$. We solve numerically $i\frac{\partial \psi}{\partial t}=H\psi$ with initial condition $\psi(x,t=0)=\left(\mathcal{U}\phi_{\epsilon}\right)(x)$ and a RHO potential. Figures \ref{campanastrana} and \ref{campanastrana2} show the resulting ``defocusing'' evolution.

\section{\label{sec:gaussian}Gaussian Function}
We examine the Gelfand triplet in Eq.(\ref{spazi_piu}) defined in sections \ref{sec:damped} and \ref{sec:rho}, in the case of the Gaussian function as element of the Hilbert space but not belonging neither to $\Phi_+$ nor to $\Phi_-$ (see figure 1). For this function, the expansion in Gamow states must be truncated and completed by an additional background function, not decaying exponentially, as discussed in sec. \ref{subs:uv}.  We illustrate theoretically and numerically the properties of the background function, specifically studying a Gaussian function $\phi(u)=\frac{e^{-\frac{u^2}{2}}}{\sqrt[4]{\pi}}$ and its transformed $\mathcal{U}[\phi](x)$. We analyse the evolution both in $(u,v)$ and $(x,p)$ planes.

\subsection{\label{subs:bg} The Background Function}
Let us define a normalized Gaussian function
\begin{equation}
\label{gausseq}
\phi(u)=\frac{1}{\sqrt[4]{\pi}}e^{-\frac{u^2}{2}}\in S(\mathbb{R}).
\end{equation}
$\phi(u)$ does not belong to $\mathcal{D}$ or to $\mathcal{Z}$ because the Fourier transformed of a Gaussian function is still a Gaussian function and $\mathcal{D}\cap\mathcal{Z}=\emptyset$.

Since $\sum_{n=0}^{+\infty}f_n^+(u)f_n^-(w)=\delta(u-w)$, we have
\begin{equation}
\label{integrale1}
\phi(u)=\int_{\mathbb{R}}dw\sum_{n=0}^{+\infty}f_n^-(u)f_n^+(w)\phi(w)\;\;\;\forall\phi\in\Phi.
\end{equation}

The integral and the summation in Eq.(\ref{integrale1}) cannot be swapped, at variance with the case of $\phi_{\epsilon}(u)$ previously considered. Therefore
\begin{equation}
\label{integrale2}
\phi(u)\neq\sum_{n=0}^{+\infty}f_n^-(u)\int_{\mathbb{R}}dwf_n^+(w)\phi(w),
\end{equation}
whence the Gaussian function has a non-trivial background. Without loss of generality, we write
$$\phi(u)=\sum_{n=0}^{N}f_n^-(u)\langle\phi|f_n^+\rangle^*+\phi_{N}^{BG}(u),$$
where $\phi_{N}^{BG}(u)$ is the N-order background.

\begin{figure}[ht!]
\begin{center}
\includegraphics[width=0.45\textwidth]{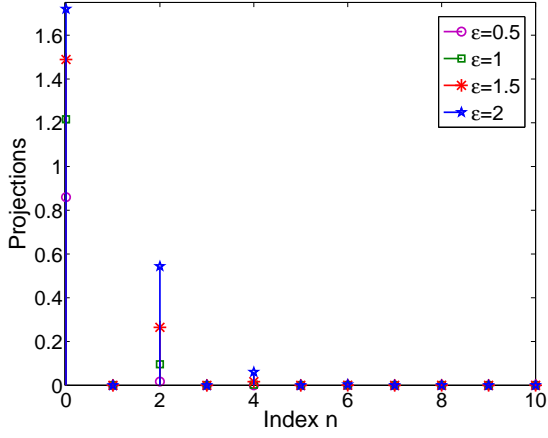}
\label{sviluppoc}
\caption{(Color online)  Projections $\langle\phi_1|f_n^+\rangle^*$ of a function with compact support [defined in Eq.(\ref{campane}), $\epsilon=1$].}
\end{center}
\end{figure}

\begin{figure}[ht!]
\begin{center}
\includegraphics[width=0.45\textwidth]{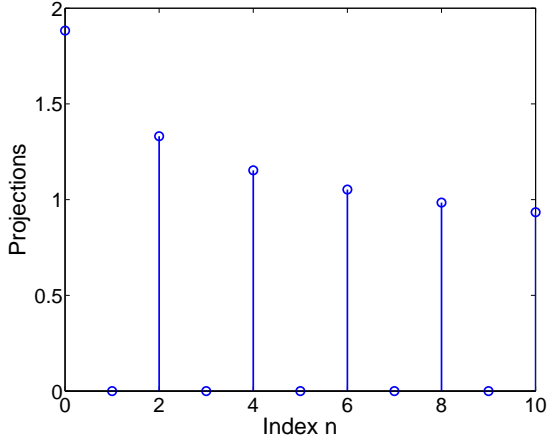}
\label{sviluppog}
\caption{(Color online)  Projections $\langle\phi|f_n^+\rangle^*$ of the Gaussian function.}
\end{center}
\end{figure}

Since $\phi(u)=\phi(-u)$, while $f_n(u)=-f_n(-u)$ for odd $n$, we have
\begin{equation}
\label{gaussian_brackets}
\langle \phi|f_n^+\rangle=\left\{\begin{array}{cc}\frac{2^{\frac{n+1}{2}}\Gamma\left(\frac{n+1}{2}\right)}{\sqrt[4]{\pi}\sqrt{n!}}&\mbox{for even } n\\0&\mbox{for odd }n\end{array}\right..
\end{equation}
Both $\langle \phi|f_n^+\rangle$ and $\langle \phi_{\epsilon}|f_n^+\rangle$ decrease for even $n$, but the Gaussian $\langle \phi|f_n^+\rangle$ decay much more slowly, as one  can see by comparing figures 9 and 10. However, this is not a mathematical proof of the existence of the background. The presence of the background is proved through the study of the initial datum: $\phi_{\epsilon}(u)$  is an initial state that is composed only by a discrete sum of resonances without any component in the continuum, because it belongs to $\mathcal{D}$; on the contrary $\phi(u)$ is an initial state with a component of continuous radiation that is the background.

We analyze the evolved N-order background wave for a Gaussian initial data. We want to study its limit as $N\rightarrow +\infty$.

We have
$$\phi_N^{BG}(u,t)=\phi(u,t)-\sum_{n=0}^Nf_n^-(u)\langle f_n^+|U(t)|\phi\rangle=$$
$$=U(t)\phi(u)-\sum_{n=0}^Nf_n^-(u)\langle f_n^+|U(t)^{\times}|\phi\rangle=$$
$$=U(t)\phi(u)-\sum_{n=0}^Ne^{-\frac{\gamma}{2}\left(2n+1\right)t}f_n^-(u)\langle f_n^+|\phi\rangle.$$
We notice that the limit as $N$ approaches infinity could diverge or not exist. This happens in most cases, and specifically for the Gaussian function. In fact, in Eq.(\ref{gaussian_brackets}), we can approximate the Gamma function
\begin{equation}
\label{gamma_stirling}
\Gamma(z)=\sqrt{2\pi}z^{z-\frac{1}{2}}e^{-z}\left[1+O\left(\frac{1}{z}\right)\right]
\end{equation}
and the factorial
\begin{equation}
\label{factorial_stirling}
n!=\sqrt{2\pi}n^{n+\frac{1}{2}}e^{-n}\left[1+O\left(\frac{1}{n}\right)\right]
\end{equation}
for large values of $z$ and $n$, thanks to the Stirling's formula \cite{gradshteyn}. We find, for even $n$
$$\langle \phi|f_n^+\rangle=\frac{2^{\frac{n+1}{2}}\Gamma\left(\frac{n+1}{2}\right)}{\sqrt[4]{\pi}\sqrt{n!}}\simeq\frac{2^{\frac{3}{4}}\left(1+\frac{1}{n}\right)^{\frac{n}{2}}}{\sqrt[4]{e^2n}}\simeq\frac{2^{\frac{3}{4}}}{\sqrt[4]{n}},$$
hence $\langle \phi|f_n^+\rangle$ approaches zero with order $\frac{1}{4}$, too slowly to let the series converge $\forall t\geq0$, $\forall u\in\mathbb{R}$, so the limit $N\rightarrow +\infty$ does not exist globally. 

This confirms that an expansion like Eq.(\ref{campanatrans}) with an infinite number of GV is meaningless for a Gaussian function, and a background term is needed.

\subsection{\label{subs:evolution} Evolution}
\begin{figure}[ht!]
\begin{center}
\includegraphics[width=0.45\textwidth]{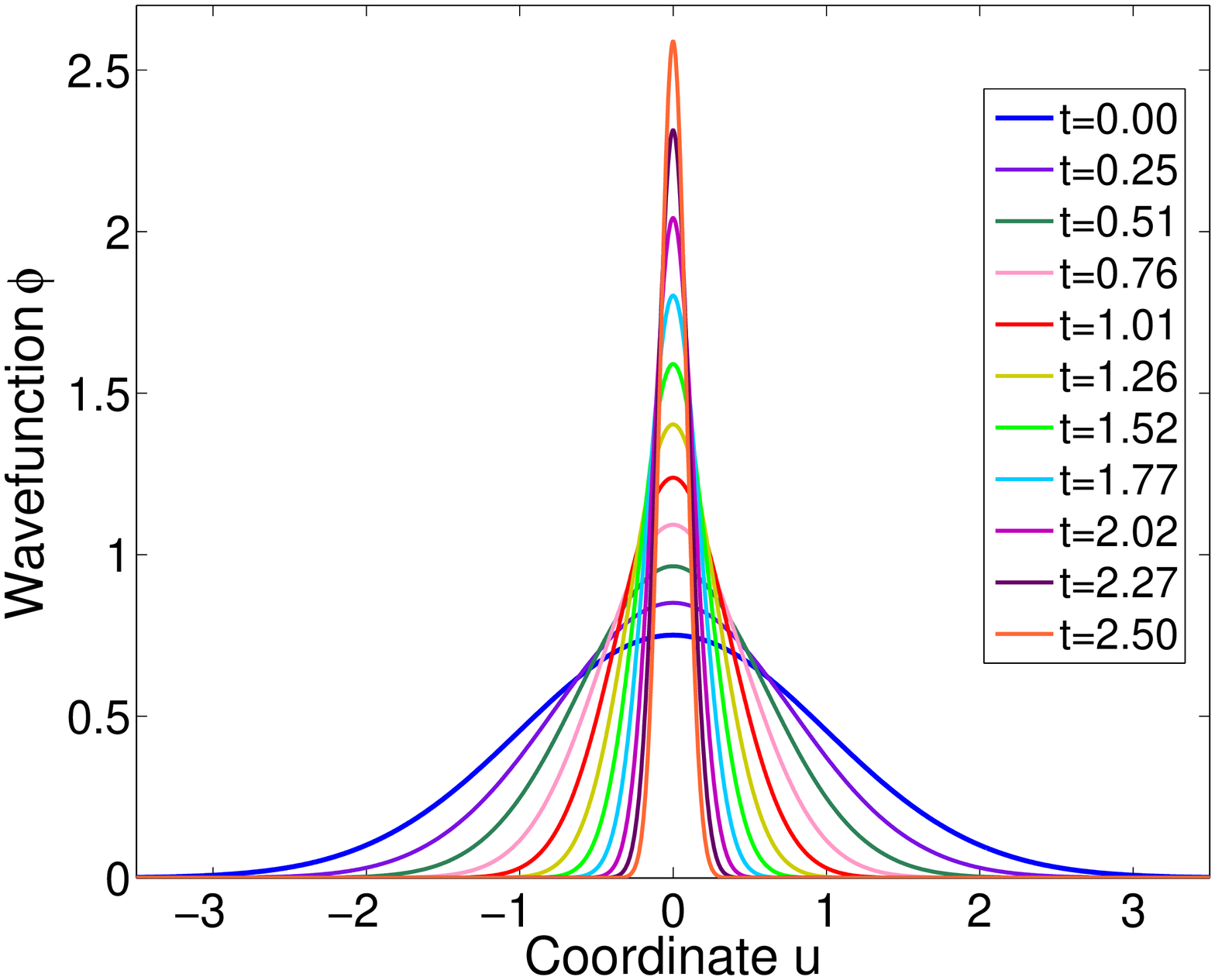}
\caption{(Color online)  One-dimensional evolution of the Gaussian wave  $\phi(u)=\frac{e^{-\frac{u^2}{2}}}{\sqrt[4]{\pi}}$.}
\label{gaussianuv}
\end{center}
\end{figure}

\begin{figure}[ht!]
\begin{center}
\includegraphics[width=0.45\textwidth]{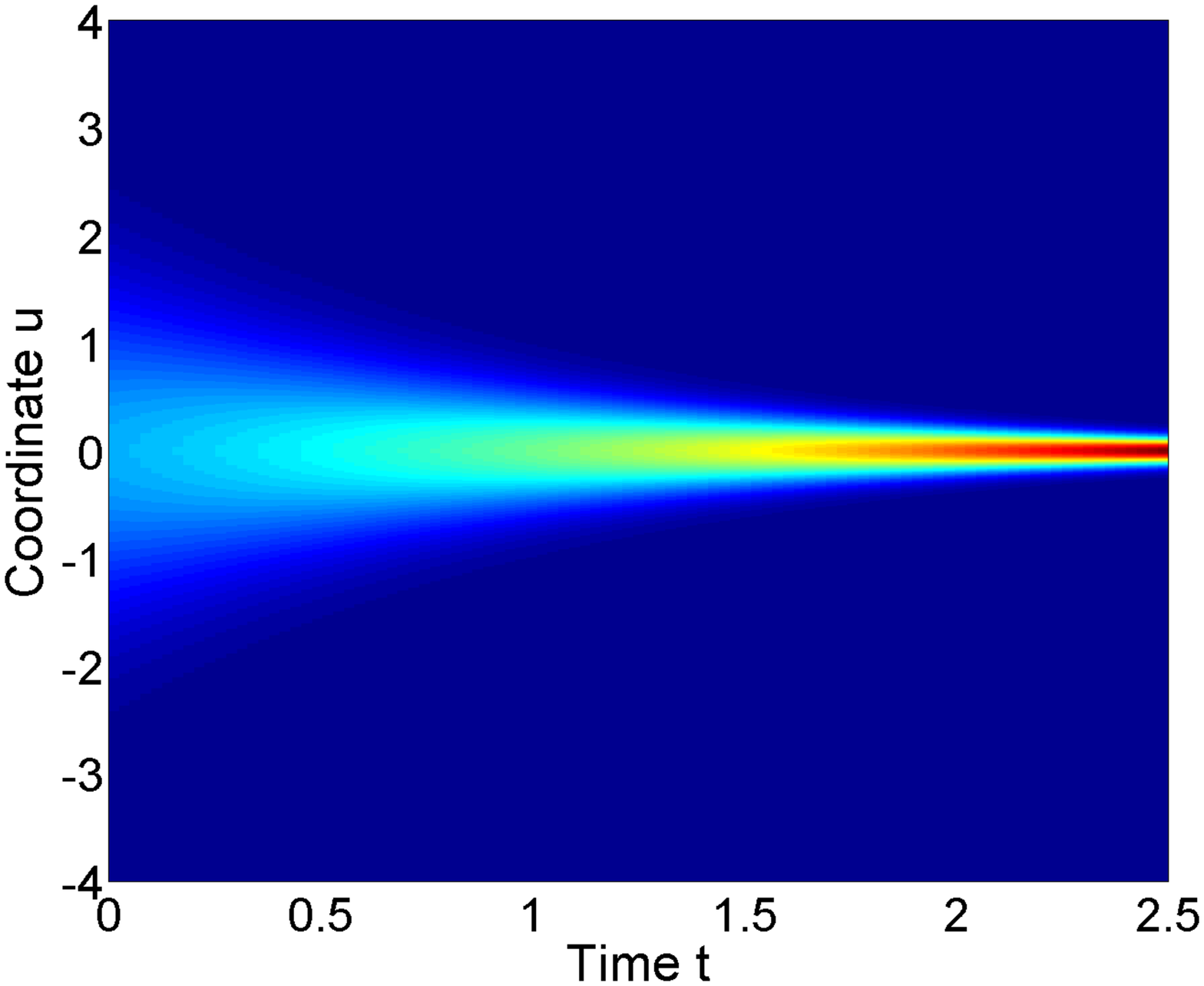}
\caption{(Color online)  Evolution of the Gaussian wave $\phi(u)=\frac{e^{-\frac{u^2}{2}}}{\sqrt[4]{\pi}}$.}
\label{gaussianuv2}
\end{center}
\end{figure}

\begin{figure}[ht!]
\begin{center}
\includegraphics[width=0.45\textwidth]{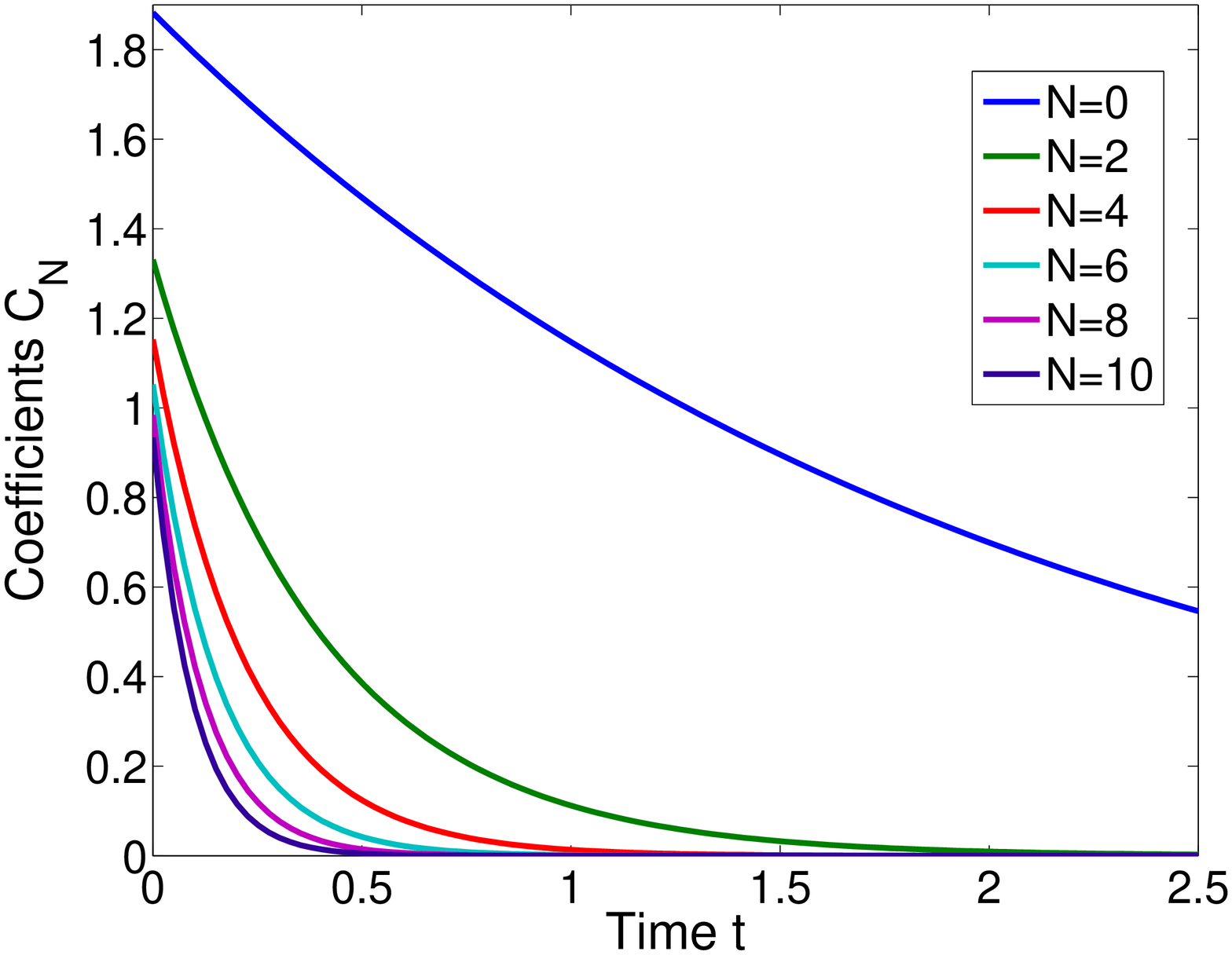}
\caption{(Color online)  Evolution of the projections $C_N(t):=\langle U(t)\phi|f_N^+\rangle^*$ on the N order resonances of the Gaussian wave  $\phi(u)=\frac{e^{-\frac{u^2}{2}}}{\sqrt[4]{\pi}}$ in the (u,v) representation, in a linear scale.}
\label{gaussiancoeffuv}
\end{center}
\end{figure}

\begin{figure}[ht!]
\begin{center}
\includegraphics[width=0.45\textwidth]{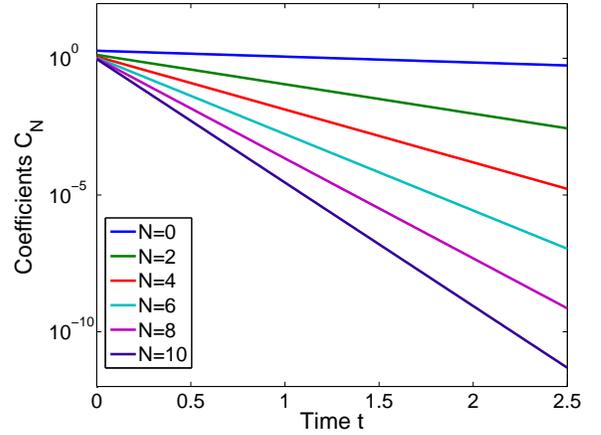}
\caption{(Color online)  Evolution of the projections on the N order resonances $C_N(t):=\langle U(t)\phi|f_N^+\rangle^*$ of the Gaussian wave  $\phi(u)=\frac{e^{-\frac{u^2}{2}}}{\sqrt[4]{\pi}}$ in the (u,v) representation, in a semilogarithmic scale.}
\label{gaussiancoeffuv2}
\end{center}
\end{figure}

Figures \ref{gaussianuv} and \ref{gaussianuv2} show a portrayal of the Gaussian function evolution. In Fig. \ref{gaussiancoeffuv} and in Fig. \ref{gaussiancoeffuv2} one can observe the decay of the coefficients.

\begin{figure}[ht!]
\begin{center}
\includegraphics[width=0.45\textwidth]{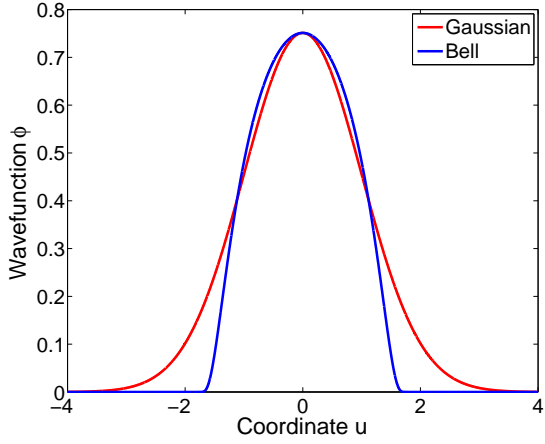}
\caption{(Color online)  Fit of the Gaussian function $\phi(u)=\frac{e^{-\frac{u^2}{2}}}{\sqrt[4]{\pi}}$ by the function $\phi_{\epsilon_0}(u)$ ($\epsilon_0=1.802425$).}
\label{confronto1}
\end{center}
\end{figure}

\begin{figure}[ht!]
\begin{center}
\includegraphics[width=0.45\textwidth]{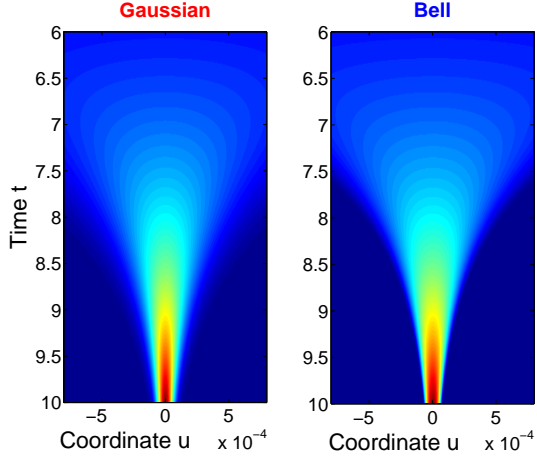}
\caption{(Color online)  Comparison of the Gaussian evolution $\phi(u,t)$ with $\phi_{\epsilon_0}(u,t)$. $\phi_{\epsilon_0}(u,t)$ focalizes without any loss or dispersion of energy, while the Gaussian presents a dispersive background (see lso Fig. \ref{confronto9}).}
\label{confronto7}
\end{center}
\end{figure}

\begin{figure}[ht!]
\begin{center}
\includegraphics[width=0.45\textwidth]{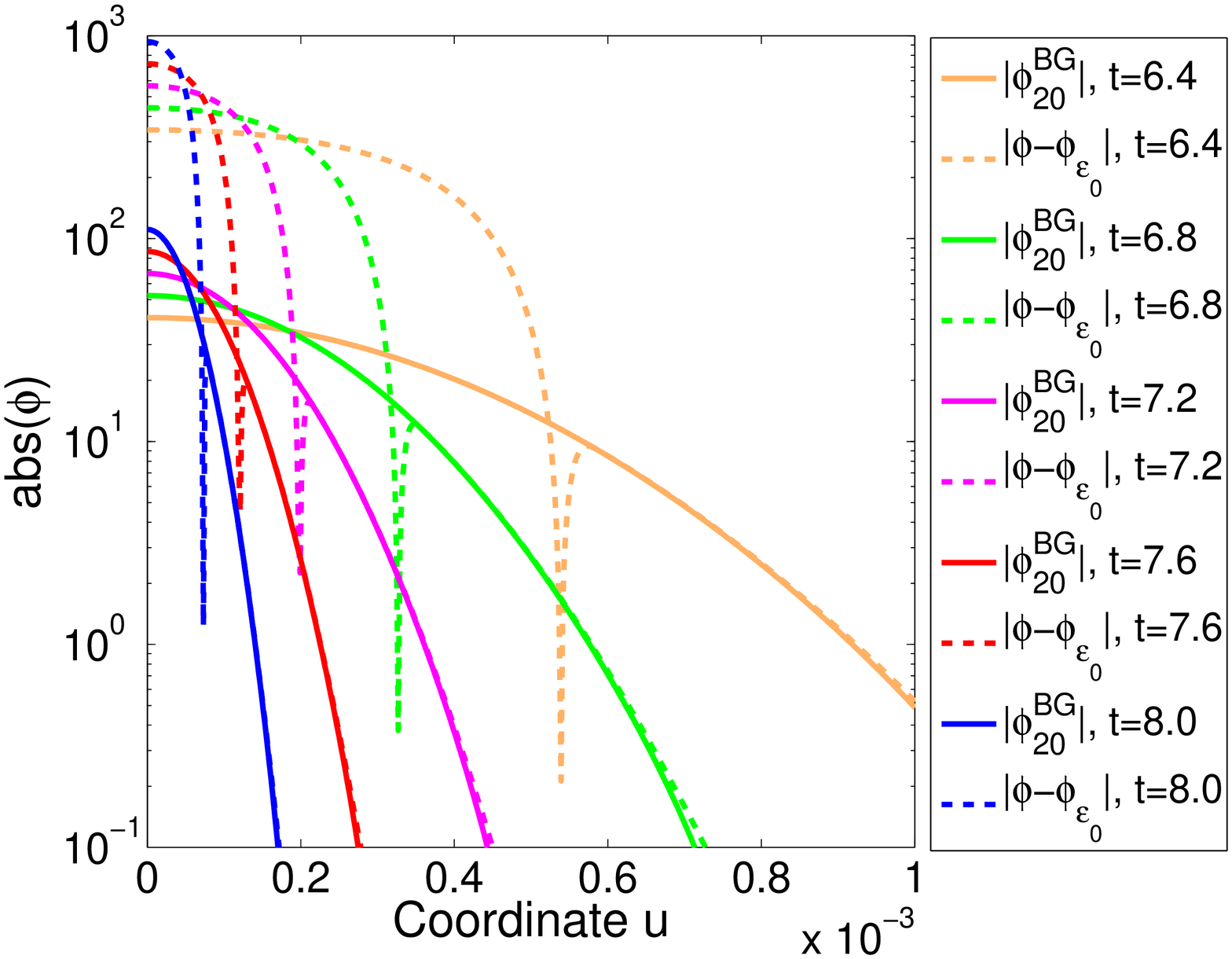}
\caption{(Color online)  We want to analyze if a region where \newline $\phi_N^{BG}(u,t)\simeq\phi(u,t)-\phi_{\epsilon_0}(u,t)$ exists. A comparison between $\phi_{20}^{BG}(u,t)$ and $\phi(u,t)-\phi_{\epsilon_0}(u,t)$ is here reported in semilogarithmic scale: these two wave packets are well overlapped on their borders.}
\label{confronto9}
\end{center}
\end{figure}

The $(u,v)$ phase space remains the simplest configuration for numerical tests of the theory. Since one runs into a high computational complexity when analyses the background evolution, we choose $(u,v)$ phase space to compare the evolution of a Gaussian function with a specific $\phi_{\epsilon}$. We best fit the normalized Gaussian function by a function $\phi_{\epsilon}$ in order to compare the background function with the difference between these two waves.
Figure \ref{confronto1} shows $\phi(u)$ and its best fit by $\phi_{\epsilon}(u)$, obtained for $\epsilon=\epsilon_0=1,802425$. Fig. \ref{confronto7} compares the calculated evolution of $\phi(u)$ and $\phi_{\epsilon_0}$. We should see the dispersive component that occurs on the boundaries of the Gaussian evolution. However, appreciating the dispersive behaviour is difficult in a linear scale; we report a comparison between $\phi^{BG}_{20}$ and $\phi-\phi_{\epsilon_0}$ in Fig. \ref{confronto9} in a semilogarithmic scale. The continuous lines represent the Gaussian background for several time values, while the dashed lines give the difference between the Gaussian and the function with compact support. One can now see without difficulties that the outlines on the boundaries are well overlapped, so the long time evolution of a Gaussian background, that is, the dispersive tail of a function not belonging to $\Phi_+$, can be approximated to the rest between the function we are studying and an appropriately chosen function $\phi_{\epsilon}$.

\begin{figure}[ht!]
\begin{center}
\includegraphics[width=0.45\textwidth]{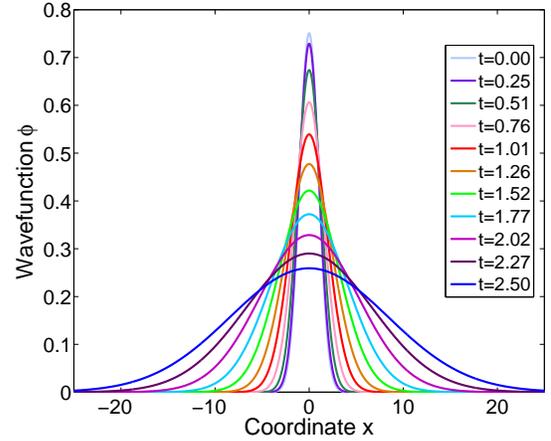}
\caption{(Color online)  One-dimensional evolution of the Gaussian wave $\phi(x)=\frac{e^{-\frac{x^2}{2}}}{\sqrt[4]{\pi}}$ under a RHO potential.}
\label{gaussiana4}
\end{center}
\end{figure}
\begin{figure}[ht!]
\begin{center}
\includegraphics[width=0.45\textwidth]{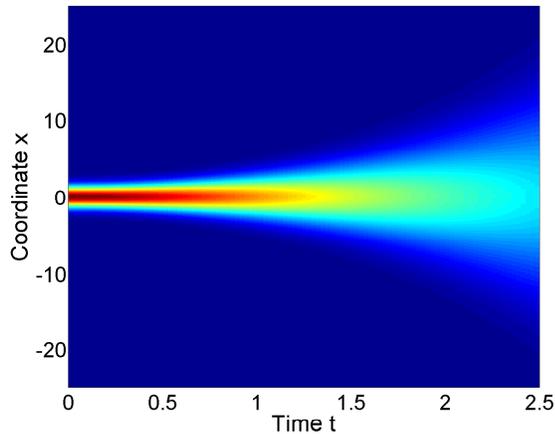}
\caption{(Color online)  Evolution of the wave $|\phi|$ with initial condition $\phi(x)=\frac{e^{-\frac{x^2}{2}}}{\sqrt[4]{\pi}}$.}
\label{gaussiana5}
\end{center}
\end{figure}

We want now complete our analysis by considering the $(x,p)-$system. By the transformation $\mathcal{U}$ we see that $\phi(x)=\left[\mathcal{U}\phi(u)\right](x)$ is a Gaussian function anyway, because
$$\left[\mathcal{U}\phi(u)\right](x)=\pi^{-\frac{1}{4}}\tilde{C}\int_{\mathbb{R}}e^{-\frac{u^2}{2}}e^{iS(x,u)}du=\left(\frac{\gamma}{\pi}\right)^{\frac{1}{4}}e^{-\frac{\gamma}{2} x^2}.$$
The focusing dynamics in the $(u,v)$ space corresponds to a defocusing propagation in the $(x,p)$ space as shown in Fig. \ref{gaussiana4} and in Fig. \ref{gaussiana5}.

\section{\label{sec:conclusions}Conclusions}
We reviewed the basic theorems and the mathematical properties underlying the time-asymmetric formulation of quantum mechanics with specific reference to the reversed harmonic oscillator.
We considered the propagation of a wave packet in the reversed harmonic oscillator within the rigged Hilbert space framework.
We specifically analysed the evolution of a class of functions with compact support in the $(u,v)$ phase space and the evolution of a normalized Gaussian function. 
For the functions with compact support we discussed the way the dynamics in the $(u,v)$ maps into the real $(x,p)$ space, and 
verify that the projections of a wave packet on Gamow states decay exponentially.
We studied the mechanism of excitation of the background function for a Gaussian function, that does not belong to $\Phi_+$. The Gaussian function cannot be expressed as an infinite linear combination of GV, 
and the results is the excitation of a dispersive wave which does not decay exponentially.

In other words, for a RHO the temporal evolution is dominated by a sum of exponentially decaying states with quantized decay rates. 
Depending on the function class of the initial conditions, one can also observe the excitation of a non-exponentially decaying component, denoted as the background.
These findings may be directly tested in the experiments by a proper shaping of the initial conditions.
We believe that our results address some of the known concepts of the RHS approach to the dynamics of unstable systems in a way that may find direct application in designing tests of time-asymmetric quantum physics,
in fields like Bose-Einstein condensation, superconductors and photonics.
Gamow vectors may also open novel possibilities in studying nonlinear waves and their reversibility properties from a fundamental point of view.

\section{Acknowledgments}
We acknowledge fruitful discussions with M. C. Braidotti, M. Materassi, P. M. Santini and P. Verrucchi. This publication was made possible through the support of a grant from  the John Templeton Foundation (grant number 58277). The opinions expressed in this publication are those of the authors and do not necessarily reflect the views of the John Templeton Foundation. 

%\newpage %Just because of unusual number of tables stacked at end
%\bibliographystyle{ieeetr}
%\bibliography{references}% Produces the bibliography via BibTeX.
\end{document}